\documentclass[aps, pre, reprint, twocolumn, superscriptaddress, showpacs, amsmath, amssymb]{revtex4-1}

\usepackage{graphicx}% Include figure files
\usepackage{dcolumn}% Align table columns on decimal point
\usepackage{bm}% bold math
\usepackage{setspace}
\usepackage{amsmath}
\usepackage{siunitx}
\usepackage{color}
\newcommand*\diff{\mathop{}\!\mathrm{d}}
\begin{document}
	\label{2.main}
	%\doublespacing
	
	%\preprint{APS/123-QED}
	
	\title{The mechanical influence of single fibre inclusions in discrete fibre networks}% Force line breaks with \\
	
	\author{DL Humphries}
	\email{Daniel.Humphries@maths.ox.ac.uk}
	\author{JA Grogan}%
	\author{EA Gaffney}
	
	\affiliation{%
		Wolfson Centre for Mathematical Biology\\
		University of Oxford 
	}%
	
	\date{\today}% It is always \today, today,
	
	\begin{abstract}{
			Semiflexible biopolymer networks are commonly found in biological systems, from the cytoskeleton of cells to the extracellular matrix. Such networks often naturally occur as composites, in which various components interact to generate rich mechanical behaviours. In this work we examine the mechanics of composites formed when a single fibre inclusion is placed within discrete fibre networks of two distinct architectures. In particular, we computationally and theoretically investigate the mechanics of composites formed when an inclusion is introduced to Voronoi- and Mikado--type networks within the nonaffine regime. On subjecting these single inclusion composites to small shear deformations, we observe different behaviours dependent on the choice of network geometry. This divergence in mechanical responses is interpreted as a consequence of architecture-dependent differences in the nonaffine displacement field. Scaling laws for the increase in network energy due to inclusions of different lengths, orientations, and elastic properties are proposed and computationally verified. We present theoretical predictions for critical values of the inclusion bending and stretching stiffnesses, above which no further increases in network energy are observed for a given shear deformation. These predictions are supported by extensive computational evidence. We expect the architectural differences identified in this work to be pertinent to theoretical investigations into the complex behaviour of biopolymer networks, and to experimental work, where complex mechanics has been observed in composite fibre networks.}
	\end{abstract}
	
	\maketitle
	
	\section{\label{sec:level1}Introduction}
	
	The mechanics of fibre assemblies has been subject to growing interest in recent decades. Materials constructed from interwoven, entangled or cross-linked fibres include synthetic examples such as paper and textiles, as well as important biological systems, such as both extra- and intracellular matrices. Common to these materials is the importance of a discrete microstructure, in which filaments organise into a complex architecture, granting the bulk material remarkable properties \citep{heussinger2007role, janmey2007negative, liu2006fibrin, hudson2010stiffening}. Of particular interest have been biological examples, including the cytoskeleton of living cells, an active and dynamic fibre network that grants cells their mechanical rigidity \citep{chugh2017actin, mizuno2007nonequilibrium}, and the hierarchical, fibrous extracellular matrix of living tissue, which interacts with many cell types \citep{licup2015stress, baker2015cell, abhilash2014remodeling}. In both of these examples, it is not sufficient to describe the system according to properties relating to the individual biopolymers, nor bulk quantities such as density. Instead, the detailed assembly of the fibres into a specific network geometry must be considered, as well as a wealth of other factors, such as cross-linking agents, motor proteins, glycosaminoglycans and liquid phases \citep{alvarado2013molecular, das2012redundancy, mattson2017glycosaminoglycans}. Many biopolymer systems are formed from semiflexible fibres, where the backbone stiffness of the fibres contributes to longer thermal persistence lengths, yet where the filaments remain far more compliant in bending than in stretching \citep{broedersz2014modeling,van2008models}. Composite networks formed from distinct species of semiflexible polymers often interact to grant the overall system diverse properties. For example, within cells, stiff microtubules interact with actomyosin, with important implications for the stiffness and strength of the fibre assembly \citep{wang2015composite, jensen2014emergent}. In extracellular matrices, collagen filaments exist across a large range of radii and organisations, from random and disordered to aligned and stiff, and interact with elastin and other polymers \citep{fratzl2008collagen, sophia2009basic, pang2017three}. Furthermore, biopolymer networks seeded with mechanically active cells form a composite system in which the stiffness of the material is determined by the biopolymer scaffold, the reorganisation of the network geometry by the resident cells, as well as the alignment and force generation of those cells \cite{jones2015micromechanics, humphries2017mechanical}. Given the remarkable properties observed in these heterotypic networks, and how frequently they occur in Nature, there is considerable interest in gaining a better understanding of the relevant physics and mechanics.
	
	Composite materials have been a topic of considerable interest in the field of continuum elasticity \citep{cates1984linear, zhang2013cross}. However, valuable insights can often be gained by considering the explicit microstructure of the system. Consider for example an individual stiff rod fixed within a simple elastic solid \citep{das2011mechanics}. In the absence of this inclusion, the displacement field is precisely affine, such that if the rod has an effect, it can only act to disrupt the affine deformation. In discrete fibre composites, the displacement field of the matrix will depend sensitively on the detailed properties of the fibre network, including its geometry and mechanical attributes \cite{heussinger2006floppy, head2003distinct}. It is not clear \textit{a priori} whether the motions of the inclusion will induce bending or stretching deformations in this microstructure, nor how additional constraints placed upon a nonaffine matrix by such an inclusion will alter the network mechanics. By adopting a discrete approach, the energy partition in individual matrix fibres, the deformation modes that are most favourable, and the local displacements of network cross-links can all be investigated. In this manner, one can gain an understanding of the bulk material behaviour from a micromechanical perspective. In turn, this readily allows for extensions, parameter variations and potentially motivates further continuum modelling of composite systems.
	
	Recent work within the context of theoretical studies into discrete fibre composites have highlighted some of the complexity inherent to these systems. \citet{van2017criticality} investigated composites detailed around a triangular lattice architecture, suggesting that a rich range of mechanical behaviours was possible, depending on the details of the network structure and the mechanical moduli of the polymers. Inclusions could act to dramatically stiffen the bulk network, up to a hundred fold, while for other parameter regimes the mechanical response was dominated by the flexible background matrix. Composites formed from two distinct polymer types have also been investigated in three-dimensions  \citep{huisman2010semiflexible}. In between regimes characterized through either a fully percolated inclusion network or isolated stiff inclusions, distinct nonlinear behaviour was identified, dependent on the ratio of stiff to compliant fibres, as well as the fibre persistence lengths. Exceptional stiffening observed in composite networks \textit{in vitro} has been discussed in the context of simulated Mikado networks, in which the rise in the composite's shear modulus was attributed to a percolation of `interphases' \citep{shahsavari2015exceptional}. Here, elliptical regions surrounding a stiff inclusion were found to be stretch dominated. As the length or density of the inclusions increases, the interactions of these elliptical interphases lead to a drastic increase in bulk stiffness. In parallel with experiments investigating the increase in network stiffness for microtubule-actin composites, \citet{bai2011mechanics} studied the mechanics of Mikado networks containing stiff inclusions and found an almost immediate rise in the stretching energy in these networks, despite the fact that the inclusions were prohibited from forming stress-bearing networks in the absence of the more compliant matrix fibres. Given that stiff filament density and geometric nonaffinity were strongly negatively correlated, the authors argued that the inclusions acted to locally suppress nonaffine fluctuations, redistributing these instead to regions of the network possessing lower densities of the stiff fibres. Related work into composites where assumptions such as monodispersity, simple cross-links and identical moduli are relaxed have also revealed a range of interesting behaviours \cite{ban2016softening, bai2011role, das2012redundancy}. While these investigations have highlighted the inherent complexity in fibre composites, little attention has been given to fibre networks containing far lower densities of inclusions, in which the dominant changes to matrix mechanics arises from the specific interactions between matrix and inclusion.
	
	In order to discuss the mechanics of discrete fibre composites, it is necessary to consider the deformations of semiflexible polymer networks and how they relate to the network geometry and fibre elasticity. Theoretical studies have discussed the existence of distinct elastic regimes in semiflexible polymer networks \citep{head2003deformation, wilhelm2003elasticity}. In particular, analysis has often focussed on the transition between a bending-dominated regime, in which the network stiffness scales with the biopolymer's bending rigidity $\kappa$, and one in which fibre stretching, governed by stiffness $\mu$, constitutes the main deformation mode \citep{heussinger2006floppy, head2003distinct, picu2011mechanics}. Within the bending-dominated regime, the local motions of cross-links are observed to decouple from the imposed bulk displacement, such that the deformation is \textit{non-affine} \citep{levine2004deformation}. This places such systems in contrast to simple continua, in which the motion of all points within the sample locally follows that of the whole. As the ratio of bending rigidity to the average distance between cross-links $l_c$ increases, the system transitions to a stretch-dominated regime, accompanied by a marked decrease in nonaffine displacement fluctuations. Various expressions for the length scale under which this affinity transition occurs have been discussed in the literature , where knowledge of the network architecture, fibre persistence length or fibre length $L$, polymer and cross-link density, spatial dimensions, and the bending and stretching moduli of the polymer all combine to place the system in one regime or the other \citep{head2003distinct, heussinger2006stiff, licup2016elastic, wang2017affine}.
	
	\begin{figure*}[t]
		\centering
		\includegraphics[width=0.95\textwidth]{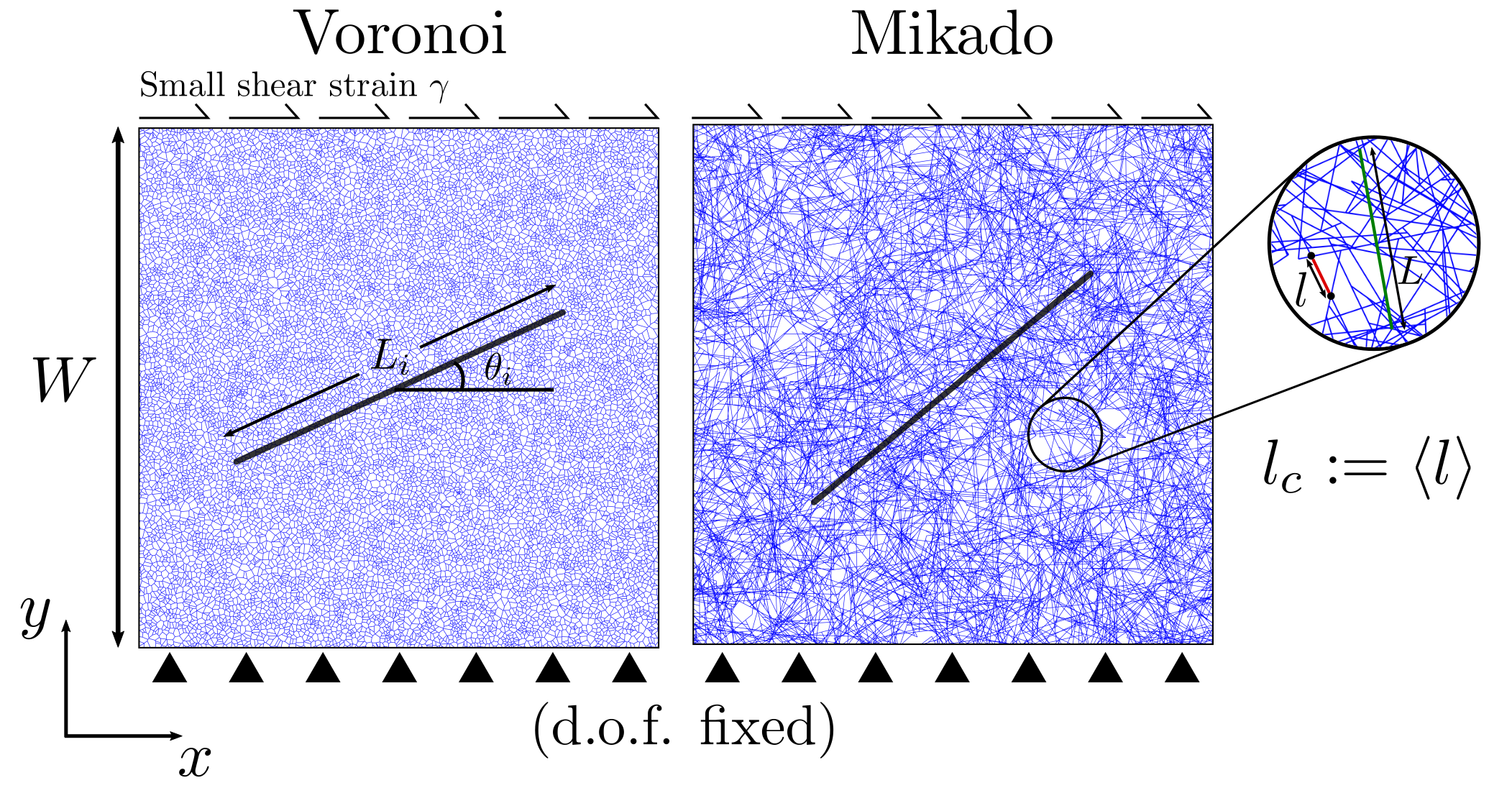}
		\caption{Illustrations of single inclusion composites used in this work. Networks are constructed within square domains of area $W^2$. Nodes attached to the lower boundary are fixed in all degrees of freedom, while those affixed to the upper boundary are shifted to produce a bulk shear of $\gamma$. Periodic boundary conditions exist on left and right boundaries. \textbf{(Left)} A Voronoi base network in which an inclusion of length $L_i$ is placed at an orientation of $\theta_i$ with the horizontal $(x)$ axis. \textbf{(Right)} A Mikado single fibre composite. The fibre extents between network cross-links $l$ take a mean value of $l_c$. Wherever fibres from the base network cross an inclusion, a new `welded' cross-link is formed. The length of Mikado base fibres is denoted by $L$. }
		\label{f0}
	\end{figure*}
	
	One can choose to characterize affinity in two different, yet related, ways. Firstly, the affine regime can be defined energetically, through the ratio of the measured bulk elastic moduli to those predicted from the energy present in fibres undergoing affine motions. Those networks for which this ratio is close to unity are deemed to be deforming affinely, while those in the nonaffine, bending-dominated regime display elastic moduli many orders of magnitude below this theoretical upper limit \citep{head2003deformation}. Alternatively, one can view the affine regime in terms of the displacement field, and its deviation from affinity \citep{levine2004deformation, wen2012non}. For example, the motion of cross-links in affine networks remains close to the theoretical prediction, while non-affine networks may involve subregions experiencing local rigid-body motions, accommodated by the bending of fibres connecting these regions \citep{bai2011mechanics}. Approaching affinity from this view is perhaps more nuanced in the following sense. As one probes different length scales within the sample, one will find locally nonaffine rotations, though these become small at length scales approaching that of the bulk sample \citep{head2003distinct, hatami2008scaling}. As such, the discussion of affinity should be mindful of additional length scales present in a problem; even rubber-like materials will display small non-affine deformations below a certain probing length scale \citep{wen2012non, wen2007local}. We expect this to be of particular importance in heterotypic networks, where additional length scales are present from multiple polymer species. In particular, one might consider the fluctuations at the length scale of one polymer to be large, yet these could be small at the scale of another, different polymer, whence the displacement field is considered approximately affine there. Furthermore, the relation between different definitions of affinity are likely to be architecture-dependent. If one considers two networks, both of which are considered equally non-affine in energetic terms, it is not clear \textit{a priori} that motions in the network will be similarly nonaffine in displacement and rotation. Indeed, while the bend--stretch transition is a generic property of subisostatic fibre networks, which can be induced by varying the network density and bending stiffness, it is nontrivial to predict the displacement behaviour of general fibre networks.
	
	We aim to investigate how properties of different network architectures can contribute to the mechanical response of composite networks. Here, we investigate how the notion of geometric fibre persistence can lead to different network mechanics for low density fibre composites. In particular, we investigate the mechanics of composites formed when single fibre inclusions are placed within fibrous (Mikado) and nonfibrous (Voronoi) architectures within the nonaffine regime, where we expect the largest deviations from continuum modelling. Different modes of deformation are identified, which depend sensitively on network architecture, and theory is developed for the scaling of network energy with inclusion length and orientation. The wide range of mechanical behaviours in these systems is also discussed, and two different transitions are identified, which are determined by the stretching and bending stiffness of the inclusion. Theoretical predictions are presented that consider the different network behaviours, and that agree well with simulation data. These results are used to generate predictions for the influence of inclusions in physical composite networks, and how their mechanics might depend on network architecture.
	
	\section{\label{sec:level2}Model}
	
	As discussed in the introduction, we shall consider throughout composites in which a single fibre inclusion is introduced into discrete fibre networks based upon two different network architectures. Namely, we shall investigate fibre assemblies constructed from Mikado \citep{latva2001rigidity, head2003deformation} and Voronoi \citep{heussinger2006stiff, heussinger2007role} geometries, which we shall refer to throughout as `fibrous' and `nonfibrous' network types, respectively. These geometries are considered as two extremes, with Voronoi networks yielding no trivial interpretation of geometric fibre persistence beyond the fibre segment, while Mikado geometries contain exactly straight fibres continuing through many cross-links. Both architectures considered in this work are two-dimensional, constructed within square domains of side length $W$.
	
	To construct Mikado networks, $N$ uniformly random points are chosen within the domain. Uniformly distributed orientations are assigned to these points. These seeds are taken as the midpoints of $N$ fibres with length $L$, giving a total line density of $\hat{\rho} = NL/W^2$. The intersections of these line segments within the domain are associated with nodes that, for large enough densities, gives rise to a connected network which spans the domain. At large enough length scales, that is for domain size $W$ chosen to be suitably large, networks are isotropic and homogeneous in fibre density and orientation. Following the identification of network nodes, or cross-links, we remove the dangling ends that remain at the end of each polymer length. This results in a network comprised of cross-links of coordination between $2$ and $4$, placing such networks below the isostatic threshold. Geometric periodicity is enforced at all boundaries, such that fibres leaving the domain along one boundary reappear at the opposite side. 
	
	\begin{figure*}[t]
		\centering
		\includegraphics[width=0.95\textwidth]{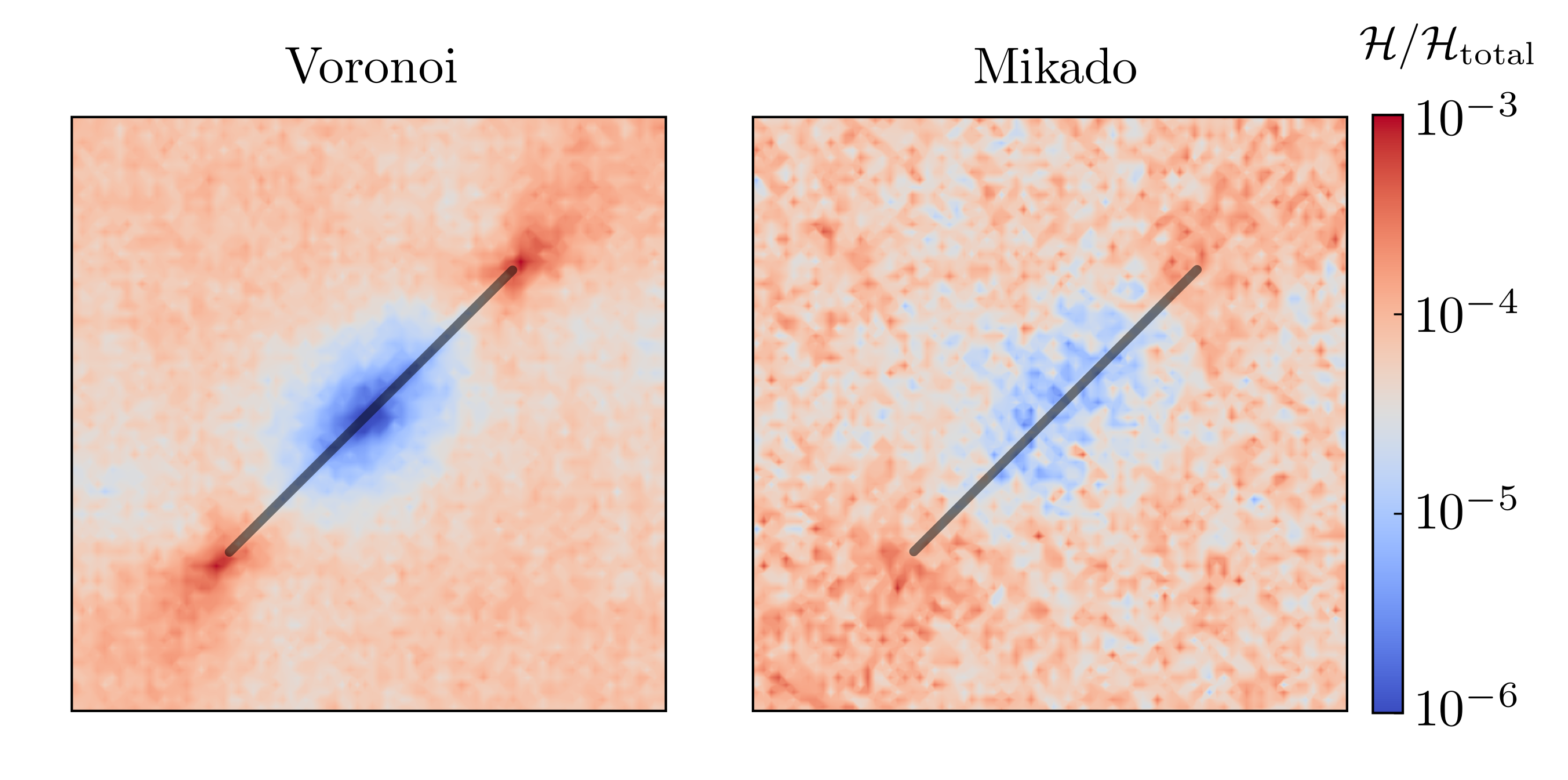}
		\caption{A visualization of the base fibre energy content $\mathcal{H}$, normalized by total network energy $\mathcal{H}_\text{total}$, in the vicinity of a single stiff inclusion. The inclusion is oriented at an angle $\theta= \pi/4$, and has length $L_i/l_c \approx 180$. When inclusions are neglected, both base networks are located within the nonaffine, bending-dominated regime, with \textbf{(Left)} Voronoi networks having $l_b/l_c \approx 10^{-3}$, and \textbf{(Right)} Mikado networks having $L/l_c = 43.5$ and $l_b/l_c \approx 10^{-4}$. We take the inclusion to be far stiffer in both stretching and bending, such that the inclusion does not experience a significant deformation, with $l_b^i / l_b = 10^{4}$ in Voronoi cases and $l_b^i/ l_b = 10^5$ in Mikado networks. In both cases, the stretching modulus for the inclusion is larger than the base modulus, $\mu_i / \mu =10^{3}$. Results are averaged over $20$ different realisations of the base network. In Voronoi networks, the effect is similar to that expected in a continuum, with base network energy reaching its largest values near the fibre end points, and an absence of energy near the centre of the fibre. Qualitatively similar but less pronounced effects are observed within Mikado composites.} 
		\label{f1}
	\end{figure*}
	
	Voronoi-type, nonfibrous matrices are generated by first seeding $M$ points within the domain, uniformly at random. This domain, complete with seeds, is then tiled into a $3\times3$ grid. A Voronoi partition is applied to the points in this extended, repeating domain, whence a geometrically periodic Voronoi diagram may be recovered by restricting the region of interest to the original square domain. Voronoi nodes, that is the vertices of the Voronoi diagram, when generated from random seeds, have node coordination $3$ in the absence of perfect symmetries. These nodes are taken to represent permanent network cross-links, and the edges of the Voronoi diagram are associated with polymer segments. Again, when considered at sufficiently large length scales Voronoi networks are statistically isotropic in density and orientation.
	
	The above procedures generate the compliant, nonaffine matrices, which we term `base' networks, into which stiff inclusions are introduced. Composites are formed by introducing a straight fibre, of length $L_i$, into a base network of a given type, where subscript $i$ refers to inclusion parameters. In these cases, the fibre midpoint is always placed in the centre of the domain. The orientation of this inclusion, $\theta_i$ is measured as a positive rotation from the $x$-axis, which is taken to be horizontal. Where this inclusion crosses base fibres, permanent network nodes are created; these are to be considered as welded cross-links, as motivated further below. Once all inclusion-base intersections have been identified, any remaining dangling ends are removed. Example networks of both types are displayed in Fig.~\ref{f0}, with Voronoi \textbf{(Left)} and Mikado types \textbf{(Right)}.
	
	We model constituent network fibres of both types, that is base and inclusion, as shear-flexible beams in the small strain regime. In particular, the internal strain energy for a generic model fibre is defined as:
	\begin{equation}
	\begin{split}
	\mathcal{H}= \frac{1}{2} \int_{L} \diff s \ \underbrace{EI \ \left(\frac{d\psi(s)}{ds} \right) ^2}_{\mathcal{H}_\text{bend}} + \underbrace{EA \ \left(\frac{du(s)}{ds} \right) ^2}_{\mathcal{H}_\text{stretch}} \\+ \underbrace{KG_{f}A \ \left(\frac{dv(s)}{ds} - \psi(s) \right) ^2}_{\mathcal{H}_\text{shear}} \ ,\label{eq_beam}
	\end{split}
	\end{equation}
	\noindent for fibre elastic and shear moduli $E$ and $G_f$, area $A$, and second moment of area $I$. The displacements along a fibre in the axial and transverse directions are given by $u$ and $v$, at fibre positions located by arc-length $s$. Shear strain is given as the difference of the plane rotation normal to the beam axis $\psi$, and the rotation of the beam cross-section $\frac{dv}{ds}$, these quantities being equal in Euler-Bernoulli beams. The fibre axial strain is $\frac{du}{ds}$ for displacement $u$ and fibre arc-length $s$. The shear correction factor accounting for the fibre cross-section is given by $K$, and the integral acts over the fibre or segment length. The internal energy may be partitioned into three distinct modes, as indicated; bending, shearing and stretching terms. Shear terms were found to be small and are inconsequential for the present study, such that we do not discuss them in the results of this manuscript.
	
	Rather than controlling the elastic fibre behaviours by varying the fibre radius and considering the cross-sectional shape, we instead employ general stretching and bending stiffnesses, $\mu$ and $\kappa$ respectively, which nominally relate as $\mu := EA, \ \kappa := EI$. However, we do not expect the mechanics of biopolymers to correspond precisely with that in classical beam physics, as it may arise from differing mechanisms depending on the polymer type, structure and scale considered \citep{fratzl2008collagen, piechocka2010structural}. As such, and in the interests of generality, we therefore view the mechanics of the system through the ratio $\kappa/\mu$ of these general rigidities, and do not return to the notion of fibre cross-sections further. By considering beam moments, we can view fibres, or fibre segments, as possessing a transverse spring constant, given as $k_\perp = 3\kappa/l^3$ for bending rigidity $\kappa$ and segment length $l$. Networks were found to be largely insensitive to the choice of fibre shear modulus $G_f$ in the range $ \left(\frac{E}{3}, \frac{E}{2} \right)$, corresponding to a Poisson ratio of between $0$ and $0.5$. Within this work, $G_f$ is taken to be half the corresponding elastic modulus, and the shear correction factor $K$ is set to unity; this has been verified to have no effect on the conclusions of this study.
	
	Summing over all fibres in the composite, as denoted by $\sum_N$, the total network energy is given by:
	\begin{equation}
	\begin{split}
	\mathcal{H}_\text{total} :=
	\frac{1}{2} \sum_N \int \text{ds} \  \kappa \left(\frac{d\psi (s)}{ds}\right)^2 + \mu \left(\frac{du(s)}{ds} \right)^2 \\+ \frac{1}{2}\mu \left(\frac{dv(s)}{ds}  - \psi(s)\right)^2 \ ,  
	\end{split}
	\label{eq_1}
	\end{equation}
	
	When considering discrete fibre networks, one may choose between different behaviours for the network cross-links. In Mikado networks, cross-links have most often been modelled as rotating joints, where crossing fibres may rotate around the cross-link without energy cost, yet translational motion in one fibre will incur bending in the other \citep{head2003deformation, picu2011mechanics}. However, in Voronoi networks there is no natural notion of geometric fibre persistence through a cross-link, given the general order $3$ node coordination. To ensure mechanical stability, cross-links are therefore modelled as fully welded in Voronoi networks, where both the translation and rotation of fibre segments at cross-links are constrained. In order to avoid constitutive differences between the networks, we also employ welded cross-links in the Mikado networks considered, as well as for those cross-links existing between inclusions and base fibres, though the differences between welded and rotating cross-links are expected to be small \citep{shahsavari2012model}.
	
	\begin{figure*}[t]
		\centering
		\includegraphics[width=0.95\textwidth]{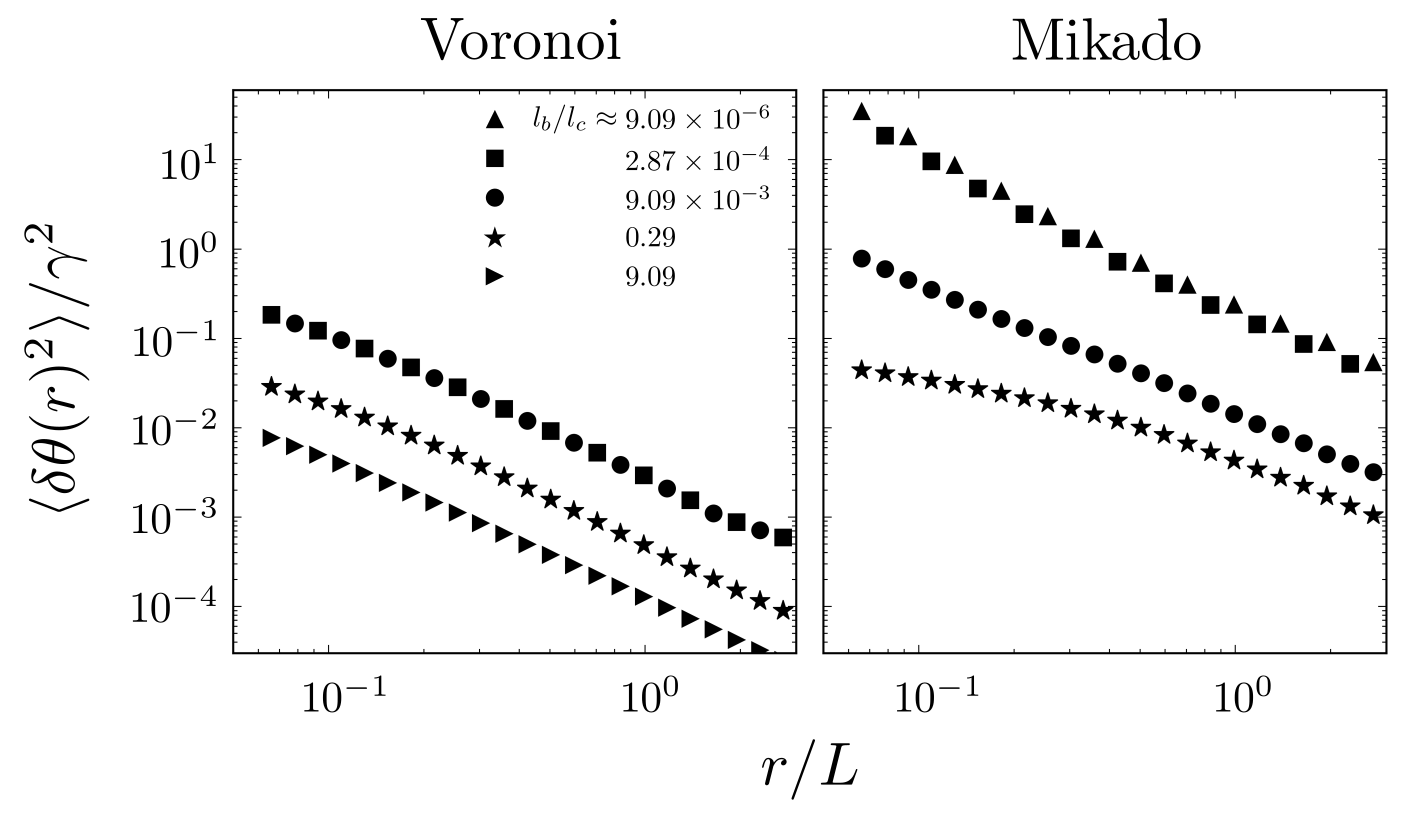}
		\caption{The angular measure of affinity $\langle \delta\theta(r)^2 \rangle := \langle\left( \theta-\theta_\text{affine}\right)^2\rangle$, which quantifies the difference between the rotation of a line formed between two sample locations a distance $r$ apart and the corresponding affine prediction, is plotted for Voronoi and Mikado base networks in the absence of an inclusion. \textbf{(Left)} In Voronoi base networks the affinity measure remains small, even for networks that are classified as highly nonaffine in terms of bending-dominated energy content. \textbf{(Right)} The same quantity is measured in Mikado composites, where we observe that for nonaffine networks (triangle and square markers) the degree of nonaffinity is more than two orders of magnitude larger than the corresponding Voronoi cases. In particular, even at larger probing distances $r/L \approx 1$, the affinity measure is larger than in the Voronoi case at the scale of the fibre segment $r/L \approx 0.06$. As such, the discussion of nonaffinity in terms of both displacement and rotation must be addressed separately to energetic definitions, as the two characterizations decouple for different network architectures. Results are averaged over $40$ realisations of the base network.} 
		\label{f2}
	\end{figure*}
	
	We simulate small bulk simple shear deformations as follows. Nodes affixed to the upper boundary are constrained in rotational and vertical displacement degrees of freedom, and are translated horizontally to give $0.001\%$ bulk shear strain, as illustrated in Fig.~\ref{f0}. Lower boundary nodes are constrained in all degrees of freedom. Periodic boundary conditions apply to nodes on vertical (left and right) boundaries in all degrees of freedom, corresponding to the geometric periodicity discussed in the context of network generation. Simulations are performed using the commercially available, implicit finite element solver Abaqus/Standard \citep{hibbett1998abaqus}. Each fibre segment, that is the fibre extent between two model cross-links, is discretized with $4$ linear shear flexible beam elements, and force balance applies to all internal finite element nodes. Beam section stiffness is derived from {Eq.~\ref{eq_beam}}. To ensure that fluctuations due to network construction remain small, the computational results presented below have been averaged over between $20$ and $80$ simulations.
	
	Having specified the details of both network generation and fibre modelling choice, we may specify the relevant geometric and mechanical parameters used within this work. Inclusions are governed by two geometric lengths; $L_i$, the total length of the straight fibre inclusion, and $l_c^i$, the mean segment length along the fibre. Mikado base networks contain the same two geometric lengths, $L$ and $l_c$, corresponding to the base Mikado fibres, with segment lengths in such networks being exponentially distributed, with probability density function $P(l_s) = e^{-l_s/l_c}/l_c$ \citep{heussinger2006stiff}. In nonfibrous networks, there is only one geometric length, the mean segment length $l_c$, to consider, owing to the lack of a geometric fibre persistence length. This mean segment length scales as the inverse of total network line density $\hat{\rho} \propto l_c^{-1}$. Finally, there is the domain length $W$, which is always taken to be sufficiently large so that size effects do not significantly affect the results presented here \citep{shahsavari2013size}. The relevant constitutive properties are as follows. Base fibres possess a bending length-scale provided by the ratio of bending and stretching stiffnesses as $l_b := (\kappa/\mu)^{1/2}$; this parameter, along with the geometric quantities discussed above determine the mechanical regime in which a base network resides. With the addition of inclusions, we also have the ratio of the stretching and bending stiffnesses for base and inclusion fibres, $\mu_i/\mu$ and $\kappa_i/\kappa$.
	
	\section{\label{sec:level3}Results}
	
	The mechanics of fibre composites are inherently more complex than those of simple continua or of monodisperse discrete fibre networks. In particular, predicting which of the competing deformation modes, including fibre bending, stretching, translations and rotations, are dominant when additional moduli and length scales are introduced into the system is nontrivial. However, if we were to consider affine base networks, where the bending stiffness of the base polymer is such that stretching deformations are more favourable, and the local strain field matches that of the bulk material, we expect to observe a picture similar to that in continuum materials. Indeed, we expect the richest behaviour to emerge within the nonaffine regime, where the base fibres exhibit large deviations from the bulk displacement field, and the networks are bending dominated. Throughout this work, it is the interplay between nonaffine base network mechanics and the additional constraints of an inclusion that we investigate. 
	
	\begin{figure*}[t]
		\centering
		\includegraphics[width=0.95\textwidth]{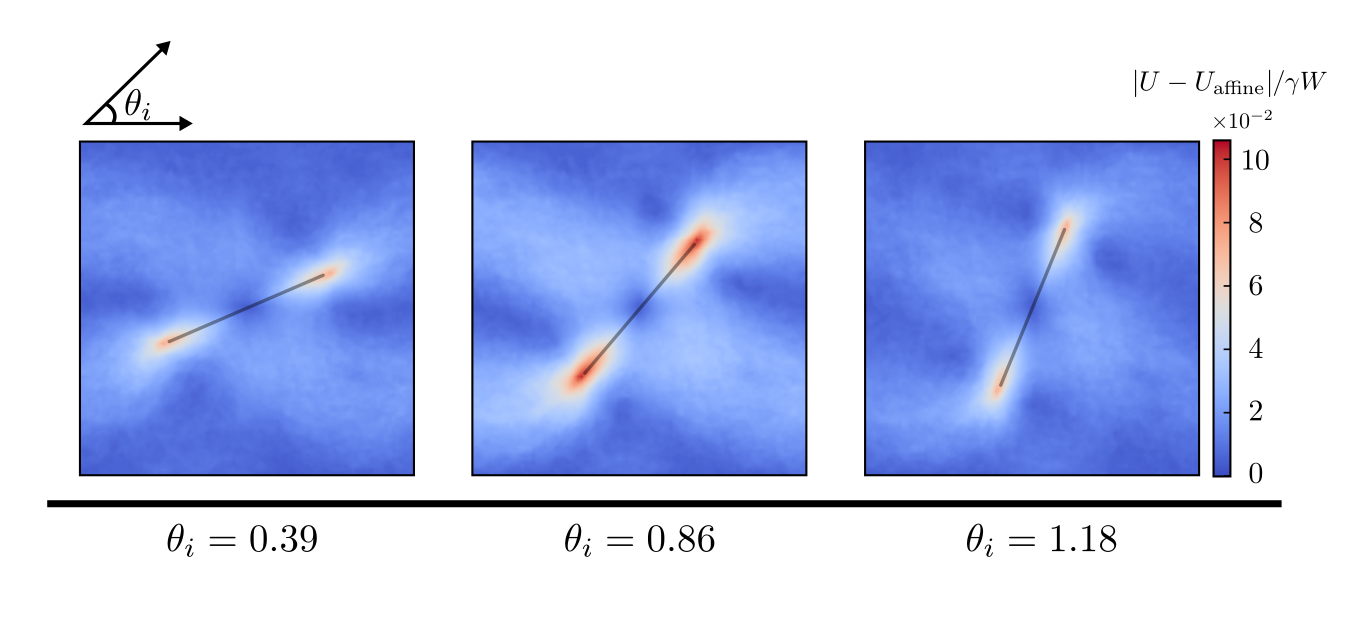}
		\caption{An illustration of the spatial deviations from the affine displacement field for network cross-links in the base network, $|U-U_\text{affine}|/\gamma W$, for stiff inclusions placed at different orientations within Voronoi networks. Inclusion and base network properties are identical to those used in Fig.~\ref{f1}. Results are averaged over $20$ realisations of the base network. Nonaffine deviations are observed to be largest for inclusion orientations approaching $\theta_i = \pi/4$, and are localized to areas close to the end points of the inclusions. The locations of highest nonaffinity correspond with regions of greater energy content, as displayed in Fig.~\ref{f1}.} 
		\label{f3}
	\end{figure*}
	
	To gain intuition into the behaviour of single fibre composites, we begin this investigation by quantifying the spatial energy profile of the base network in the presence of an inclusion. Fig.~\ref{f1} displays the energy contained within fibres, normalized by the total network energy, for base networks within the non-affine regime. The inclusion, oriented at an angle $\theta_i = \pi/4$ with the $x$-axis and possessing moduli such that it is far stiffer in both extension and bending than the base network fibres, has a pronounced effect on the spatial distribution of energy in the base network. In nonfibrous (Voronoi) matrices, despite their discrete nature, the measured result is qualitatively similar to that expected in continuum models, in which the inclusion acts to incur significant energy increases at its termination points. Base fibres along the length of the inclusion, away from the end points, experience smaller deformations than elsewhere in the matrix, resulting in a large reduction in energy in this region. In fibrous (Mikado) architectures, the comparison with the continuum system is less clear. While qualitatively similar behaviours are observed, with a slight increase in network energy located at the inclusion ends, and a reduction in fibre energy in the vicinity of the inclusion midpoint, the effect is less pronounced than in nonfibrous base networks. 
	
	We interpret this result in terms of the nonaffine displacement field. In particular, we can consider nonaffine displacement fluctuations in terms of the translations, of characteristic magnitude $d$, of floppy fibres \citep{heussinger2006floppy}. This translation, whence the magnitude, is defined at the whole fibre scale, length $L$, such that we expect no dependence on segment length $l_c$, and we are left to conclude that $d\sim L$. In nonfibrous matrices, such as the Voronoi architecture considered here, nonaffine deformations are expected be at the scale of the mean segment length $l_c$ instead, owing to the lack of an additional fibre length, which is naturally smaller than the Mikado fibre length. As such, we expect the inclusion to experience smaller nonaffine displacement fluctuations in nonfibrous architectures than in Mikado composites, even when both networks display a similar nonaffinity as quantified energetically. This prediction shall be investigated later in this work, where we explicitly examine the consequences of different nonaffine deformation fluctuations. We also hypothesize that the inclusion acts as a probing length for the size of the nonaffine deformation field. While the deviation from affine displacements in the nonfibrous cases is small at the length of the inclusion, we expect it to be larger for Mikado cases, leading to a disruption of the spatial energy profile expected in continuum composites \cite{das2011mechanics}.
	
	\begin{figure*}[t]
		\centering
		\includegraphics[width=0.95\textwidth]{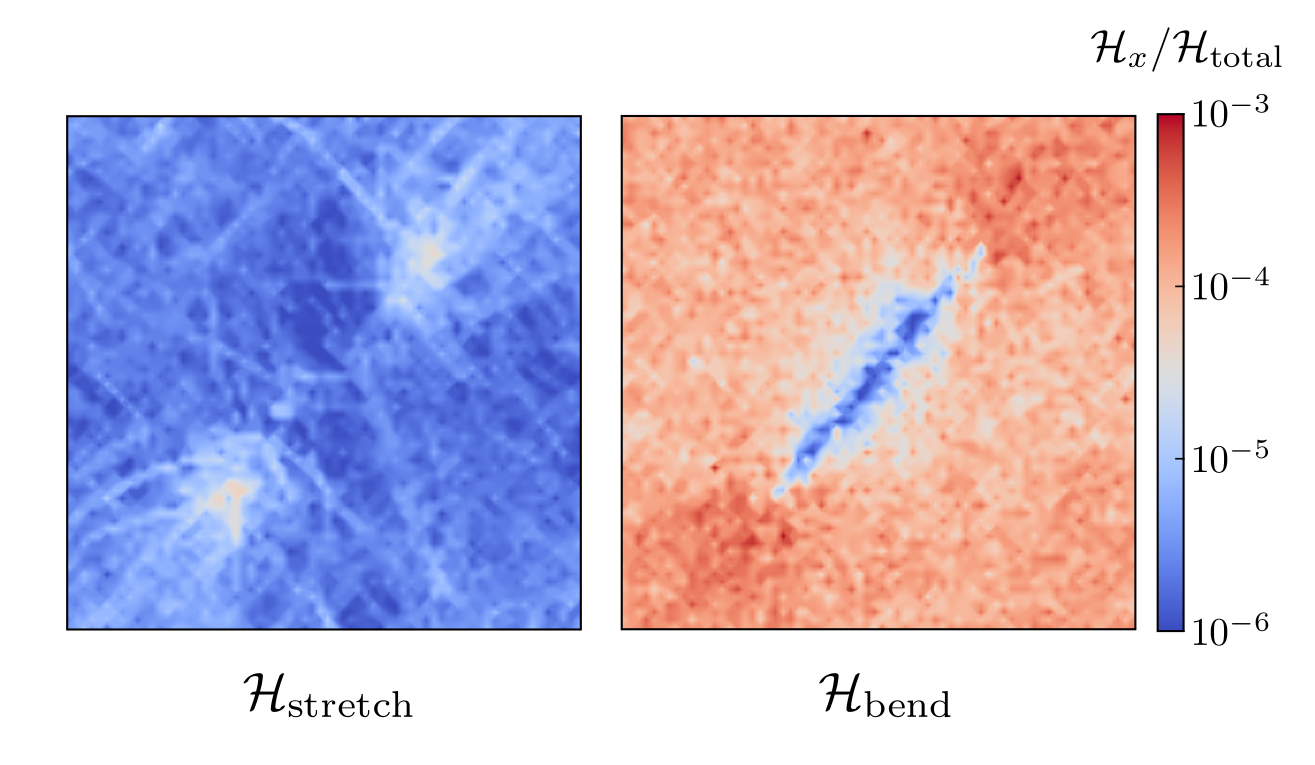}
		\caption{The spatial partition of energy in Mikado composite networks is illustrated for the same networks as in Fig.~\ref{f1}, displaying the stretching and bending contributions to the total fibre energy in base network segments. No significant increase in stretching contributions \textbf{(Left)} along the inclusion body is observed, though base networks do experience a marked decrease in bending energy \textbf{(Right)} adjacent to the inclusion. As such, we expect that the observed increase in bend-to-stretch ratio noted elsewhere is instead the consequence of a decrease in bending energy, rather than an increase in base fibre stretching.} 
		\label{f4}
	\end{figure*}
	
	One can quantify deviations in the displacement field by considering the affinity measure $\langle \delta\theta(r)^2 \rangle = \langle (\theta - \theta_\text{affine})^2\rangle$, where $\theta$ is the observed rotation, $\theta_\text{affine}$ is the corresponding angle change in a precisely affine deformation, and $r$ is the distance between two material points. This measure has been used to discuss a broad categorization of the displacement affinity of fibre networks \cite{head2003distinct}. In particular, we consider lines drawn between any model cross-links separated by a distance $r$, whence $\langle \delta\theta(r)^2\rangle$ measures the deviation from the affine rotation angle. The affinity measure for bending dominated base networks is plotted in Fig.~\ref{f2}, for varying degrees of energetic nonaffinity in the base architecture, that is for varying ratio bending length scale to mean fibre segment length $l_b / l_c$. Even for the most heavily bending-dominated Voronoi networks (\textbf{(Left)}, squares), $\langle \delta\theta(r)^2\rangle$ is still smaller for equivalent probing distances $r$ than for networks approaching the bend-stretch transition in Mikado cases (\textbf{(Right)}, circles). In particular, for networks of different architectures that possessed similar energetic nonaffinity, the deviations in rotation angles from the affine prediction are far larger in Mikado networks than in Voronoi cases. If this is related to an inclusion twice the length of a base Mikado fibre, $r/L = L_i/L = 2$, this leads to a $100$ fold larger displacement measure of nonaffinity in fibrous base networks. As such, even at large probing distances, composites with Mikado architectures exhibit significant nonaffine displacements and rotations when compared with their nonfibrous counterparts.
	
	We have discussed how the nonaffinity, as quantified through deviations in local rotations, differs depending on network architecture. In order to relate nonaffinity in the displacement field with the spatial energy profiles observed in Fig.~\ref{f1} we wish to visualise a spatial measure of nonaffinity in the presence of a stiff inclusion. In Fig.~\ref{f3}, we plot the magnitude of the difference between the displacement field $U$ and the affine prediction $U_\text{affine}$ for base network cross-links, $|U - U_\text{affine}|$, as measured for nonaffine, nonfibrous base networks and inclusions placed at three different orientations $\theta_i$. We observe that this quantity is largest at angles approaching $\pi/4$, and reduces as the angle approaches $\theta_i=0, \ \pi/2$. These nonaffine deviations occur at the same locations as the large increase in spatial energy observed in Fig.~\ref{f1}. As such, it is clear that the mechanics of such networks can be understood through considering the affine strain field, and how both the base network and the additional constraints placed upon it by the inclusion lead to departures from this prediction. 
	
	Before proceeding with scaling arguments for the energy increase when inclusions are introduced to nonaffine networks of different geometries, it is prudent to briefly discuss observations made elsewhere in the literature. In particular, similar composite networks have been investigated in other work \citep{shahsavari2015exceptional}. Here it was suggested that an inclusion, oriented at an angle $\theta=\pi/4$ with the horizontal, acted to locally increase the stretching energy in base fibres adjacent to the inclusion. This increase in peripheral stretching energy could then explain a stiffening effect in networks containing many stiff inclusions, where these `interphases' percolate at a density dictated by shape of the ellipses containing the stretching-dominated region. In order to compare with this theory, we investigate the spatial energy partitions in nonaffine Mikado single fibre composites in Fig.~\ref{f4}. The spatial energy partition for base fibres is displayed, normalized by the total network energy, and visualized with log-scale colours, emphasizing the large variations in energy content. While there is a modest increase in the stretching contributions in base fibres close to the inclusion ends, similar to that observed in Fig.~\ref{f1}, there is a drastic \textit{reduction} in base fibre bending energy along the fibre inclusion body, where the inclusion acts to prevent significant fibre deformations. As such, we posit that the stretch-dominated region observed elsewhere is due to a reduction in the bending energy, rather than significant increases in the stretching energy. We now instead proceed with a discussion of the displacement affinity in the base network and how this relates to the observed increase in network energies.
	
	\begin{figure}[t]
		\centering
		\includegraphics[width=0.46\textwidth]{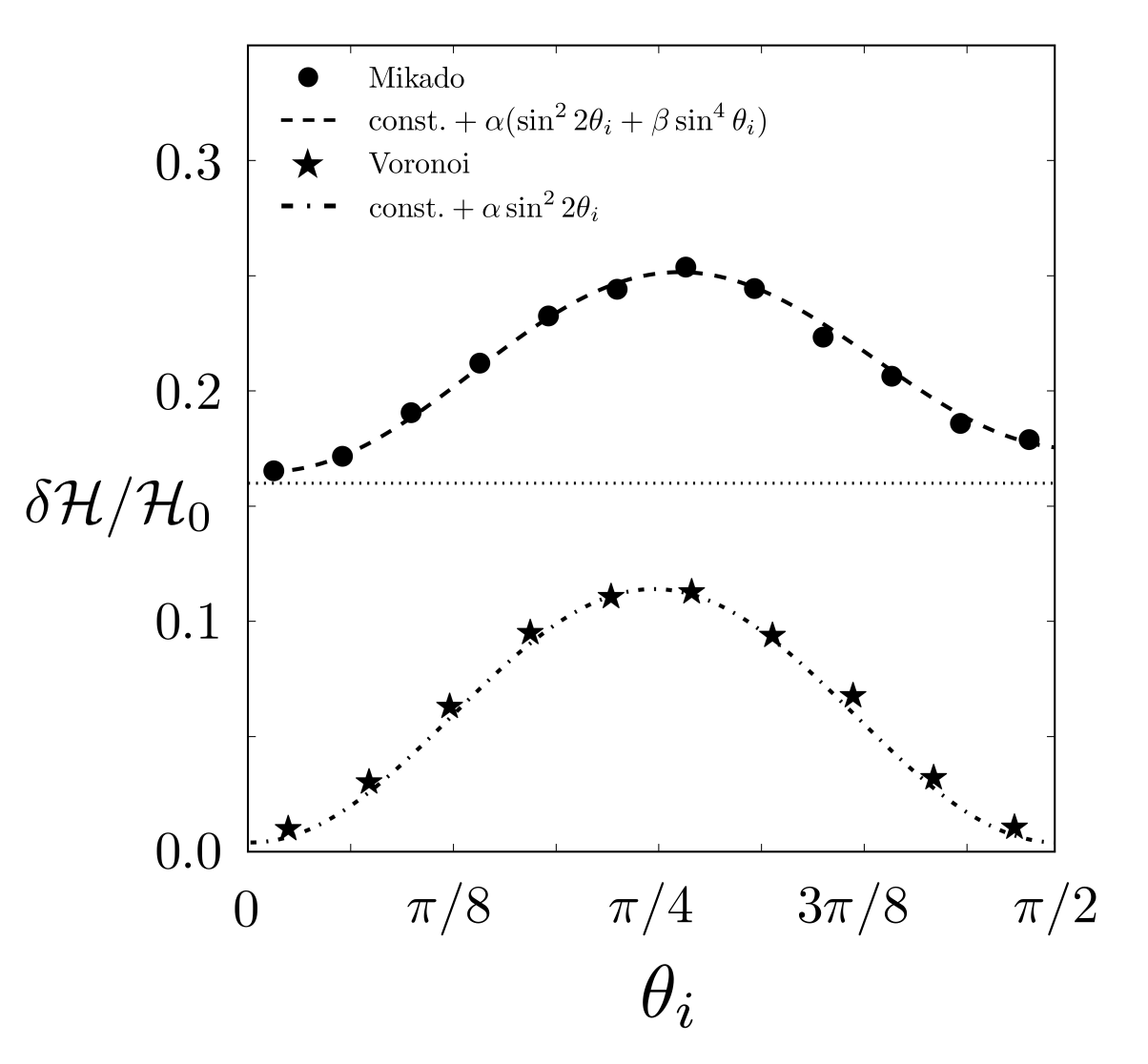}
		\caption{The percentage increase in total network energy $\delta\mathcal{H}/\mathcal{H}_0$, where $\mathcal{H}_0$ refers to the total energy content in the base network in the absence of the inclusion, is shown for varying inclusion orientation $\theta_i$. The elastic moduli for the inclusion are taken such that the inclusion is stiff in extension and bending. When Voronoi base networks (stars) are considered, the data closely fit a curve proportional to $\sin^2 2\theta_i$, corresponding to a correction to the affine strain field in the axial direction of the fibre. In Mikado composites (circles), the increase in network energy is asymmetric, with a larger increase observed for vertical ($\theta_i = \pi/2$) than for horizontal ($\theta_i=0$) fibres. Introducing a contribution accounting for the affine rotation angle, $\sin^4\theta_i$, allows for a good agreement with the observed data. In Mikado simulations, there is a notably greater increase in network energy in the presence of a stiff inclusion, even for horizontal fibres, for which both the rotational and axial contributions predicted in scaling theories vanishes, owing to the local suppression of larger nonaffine fluctuations. Results are averaged over $40$ realisations of the base network.} 
		\label{f5}
	\end{figure}
	
	We have observed that network energy contributions are largest towards the ends of fibre inclusions in both Mikado and Voronoi composites. We interpret this observation as follows. Consider an affine displacement applied to all sample points $[x,y]$ induced by a small shear $\gamma$, that is:
	\begin{equation}
	\begin{bmatrix}
	X\\
	Y
	\end{bmatrix}=\begin{bmatrix}
	1 & \gamma  \\
	0 & 1 
	\end{bmatrix}
	\begin{bmatrix}
	x \\ 
	y
	\end{bmatrix} \ .
	\end{equation}
	\noindent We expect that at the length of the inclusion, the base fibre displacement field is close to the affine prediction, especially in the nonfibrous base architectures. This affine displacement can be applied to an inclusion described by the vector $[x, y] = L_i[\cos\theta_i,\sin\theta_i]$, where $\theta_i$ determines the orientation of the inclusion with the horizontal. Accordingly, the stiff inclusion would experience an axial strain equal to $\frac{1}{2}\gamma \sin2\theta_i$, to first order in the small strain $\gamma$. This would in turn lead to a considerable energy contribution, proportional to $\mu_iL_i\sin^22\theta_i$. For sufficiently stiff inclusions, this contribution is expected to be large, such that it is more energetically favourable to relax this inclusion strain at the expense of deformations in the background, compliant matrix. This involves a displacement in the base matrix in the axial inclusion direction proportional to $\gamma L_i\sin2\theta_i$. From Fig.~\ref{f1} and Fig.~\ref{f3}, we expect the displacements and the majority of the energy contribution to be limited towards the ends of the rods, with a region close to the inclusion centre undergoing a rigid body rotation. Therefore the scaling of the increase in network energy $\delta\mathcal{H}$ should follow as:
	\begin{equation}
	\left\langle \delta\mathcal{H} \right\rangle \sim \gamma^2L_i^2\sin^22\theta_i, \label{eq_Li}
	\end{equation}
	\noindent An important aspect of this prediction is that inclusions placed perpendicular or parallel with the shearing direction should lead only to negligible increases in the network energy, as the affine strain of the inclusion vanishes to first order. The energy increase for varying fibre orientation $\theta_i$ between $\theta_i=0, \pi/2$ is quantified in Fig.~\ref{f5}. We find that the scaling prediction captures the behaviour of the nonfibrous Voronoi base networks well, with a significant percentage increase in network energy for fibres oriented around $\theta_i = \pi/4$, decaying to almost no increase for horizontal or vertical fibres. While the data for Mikado networks initially appear to follow this theory, we observe that the data curve is asymmetric in this case, and ceases to follow the $\sin^22\theta_i$ prediction for vertical fibres, that is $\theta_i = \pi/2$. This deviation and asymmetry can be explained by considering displacements transverse to the fibre inclusion axis. In particular, we return to the observation that there exists a very low energy region towards the midpoint of the inclusion; we expect that this reduction in energy occurs due to incidental base fibres undergoing rigid body rotations with the stiff fibre, as all other deformation modes would lead to an increase in fibre energy. We therefore predict that a transition region emerges, in which filaments from the nonaffine background matrix interact with those from the rigid body region. In terms of our chosen parameters $\theta_i, \ L_i, \ \gamma$, the affine rotation angle for the inclusion can be derived from the cosine rule as follows. Let the deformed inclusion length be given by $L_i^\epsilon$, then the affine rotation angle $\phi$ follows from:
	\begin{align}
	\gamma^2L_i^2\sin^2\theta_i = L_i^2 + (L_i^\epsilon)^2 - 2L_iL_i^\epsilon\cos\phi \ ,
	\end{align}
	\noindent so that for small $\gamma, \ \phi$, 
	\begin{align}
	\phi \approx \gamma \sin^2\theta_i \ .
	\end{align}
	Consider a base fibre, length $L$, that crosses the transition between the rigid body region and the background matrix. This fibre rotates according to the affine rotation $\phi$ of the inclusion, inducing (or incurring) additional displacements of order $L\phi\sim L\sin^2\theta_i$. As the number of such fibres scales as $L_i/l_c$, and the number of base fibres coincident to a given transition fibre scales as $L/l_c$, we expect the corresponding energy induced by these displacements to scale as:
	\begin{align}
	\left\langle \delta\mathcal{H}_\phi \right\rangle &\sim \frac{L_i}{l_c}\frac{L}{l_c} \left\langle\frac{\kappa_b(L\phi)^2}{l^3} \right\rangle \ , \\
	&\sim \frac{L_iL}{l_c^2}\left\langle\frac{\kappa_b(L\gamma\sin^2\theta_i)^2}{l^3} \right\rangle \\
	& \sim f(L, l_c) \kappa_b \gamma^2 L_i\sin^4\theta_i \ ,
	\end{align}
	\noindent where averaging takes place over base fibre segment lengths $l$. We return to the form of the scaling function $f$ later in this work, though the scaling with $\sin^4\theta_i$ is sufficient for the discussion here. If we consider an additional contribution to the energy change in Mikado composites, of the form $\sin^22\theta_i + \beta\sin^4\theta_i$, as shown in Fig.~\ref{f5}, dashed, we find that we can explain the asymmetry in $\theta_i$ for data from Mikado single inclusion composites and achieve a good agreement between data and theory. In both single inclusion composite types the amplitude of the data curve relative to values for horizontal or vertical inclusions is similar, as expected for the axial nonaffine displacements induced by the inclusion. We note that we do not expect to find significant asymmetry in Voronoi cases, as there is an absence of long-range correlations, enforced by long base fibres that can span both regions. Interestingly, while almost no increase in total network energy is incurred for horizontal and vertical inclusions in Voronoi base networks, a significant increase is observed for Mikado composites, even for horizontal fibres, where both of the energy contributions discussed in the above theory vanish. This may be understood by considering the larger nonaffine fluctuations in Mikado base architectures. Those fibres crossing the stiff inclusion are prevented from undergoing the nonaffine displacements that minimize energy in the base network. As such, constraints on these fibres induce deflections in other coincident base fibres, leading to a larger increase in energy, with approximately a $25\%$ increase observed for $\theta_i=\pi/4$ in Mikado composites compared with $\approx11\%$ in nonfibrous architectures. In Voronoi cases, the additional constraints only lead to small corrections owing to the far smaller nonaffine displacements in this architecture. 
	
\begin{figure}[t]
		\centering
		\includegraphics[width=0.48\textwidth]{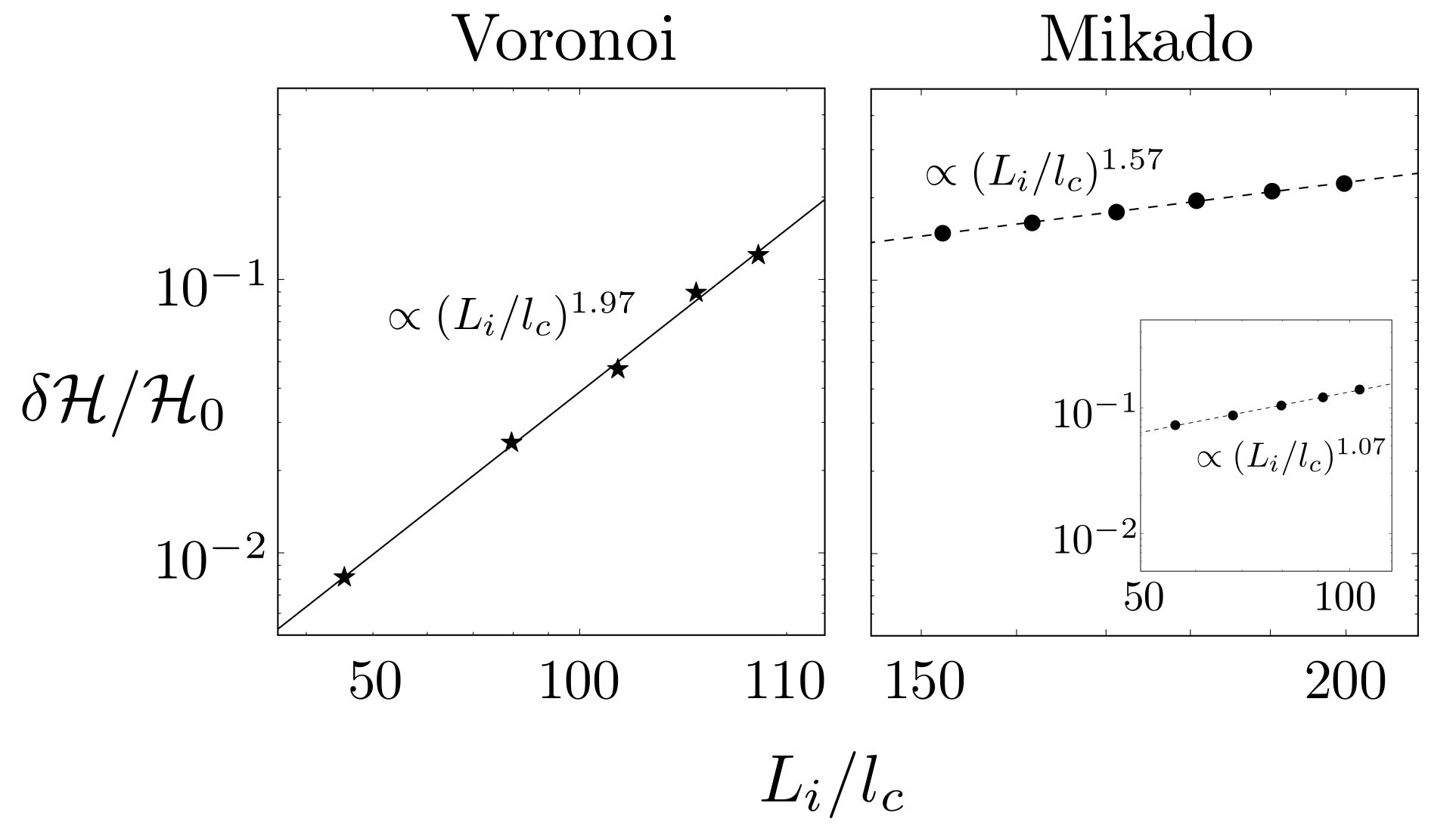}
		\caption{The scaling of the change in total network energy $\delta\mathcal{H}/\mathcal{H}_0$ with inclusion length $L_i$ is plotted, for both Voronoi \textbf{(Left)} and Mikado \textbf{(Right)} base networks within the nonaffine regime. The change in energy for non-fibrous, Voronoi base cases scales according to $L_i^2$, as expected for an axial correction to the affine displacement field by the stiff inclusion. A weaker dependence is noted for Mikado base networks. In particular, for longer inclusions $L_i/l_c > 150$, the scaling with $L_i$ is approximately equal to $L_i^{1.57}$. For shorter inclusions in Mikado networks, this scaling reduces to close to unity. Results are averaged over $40$ realisations of the base network.} 
		\label{f6}
	\end{figure}

	As discussed above, the correction to the affine displacement field in the tangent direction of the inclusion is expected to scale with the square of the length $L_i$, assuming a correction to the affine strain of the inclusion towards the fibre end points. We investigate this prediction in Fig.~\ref{f6}. When Voronoi base cases are considered \textbf{(Left)}, we find close agreement with this prediction, with network energy increase scaling as $(L_i/l_c)^{1.97}$ for constant $l_c$. However, the energy increase in Mikado networks deviates from this theoretical argument in Fig.~\ref{f6}~\textbf{(Right)}. In particular, for smaller stiff inclusions, similar to the length of the base fibres, a far weaker scaling with $L_i$ was observed, where $\delta\mathcal{H} \sim L_i^{1.07}$. When far longer inclusions were considered, the dependence on $L_i$ became more prominent, as $\delta\mathcal{H} \sim L_i^{1.57}$. We interpret this as a consequence of the scaling of nonaffine displacements in the base networks, as well as the various competing modes through which energy increases in fibrous composites. Indeed, the theory presented above breaks down if the surrounding matrix does not deform according to an affine displacement field. This is the case for short inclusions, which rotate and translate according to a locally nonaffine displacement field, and the increase in energy is largely due to the suppression of nonaffine base fibre deformations. We expect that to recover the theoretical scaling with $L_i^2$, the quantity $L_i/L$ in Mikado composites must be of similar size to $L_i/l_c$ in Voronoi cases. However, as this would involve fibre network systems larger than we are able to investigate in this work, we do not discuss this further.
	
	We have thus far considered fibre inclusions that are far stiffer with respect to both extension and bending than the base network. We now proceed to relax these assumptions and investigate how the change in network energy scales as $\mu_i$ and $\kappa_i$ increase from compliant to stiff. 
	
\begin{figure*}[t]
		\centering
		\includegraphics[width=0.95\textwidth]{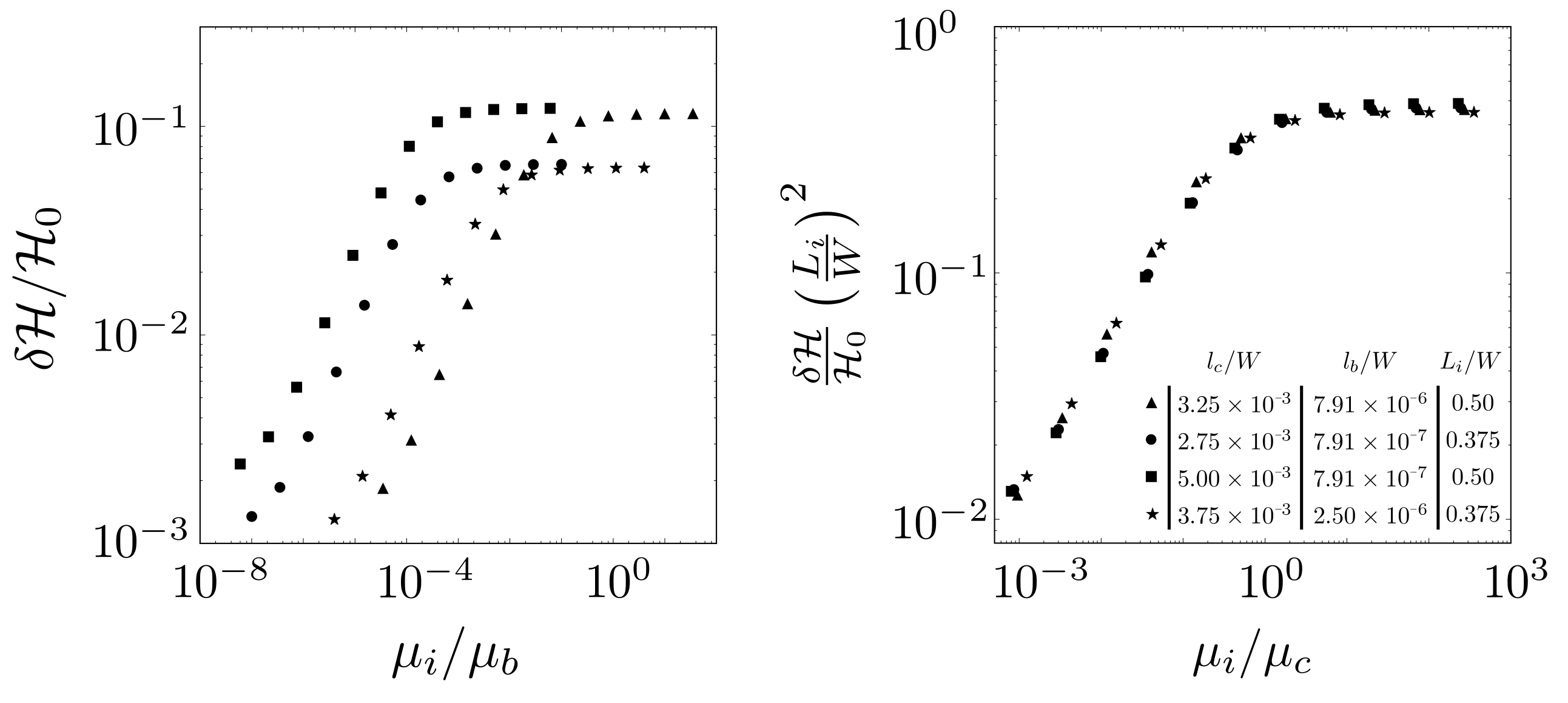}
		\caption{The percentage increase in total network energy for Voronoi composites is investigated as the inclusion stretching modulus $\mu_i$ varies. \textbf{(Left)} The increase for four different parameter choices for the base network density and bending stiffness as well as for the inclusion length are considered for varying $\mu_i$, normalized by the base fibre rigidity $\mu_b$. We note that there exists a transition, where above a critical value $\mu_c$ of the inclusion stretching stiffness, further increases in $\mu_i$ do not lead to likewise rises in total network energy. \textbf{(Right)} The same results are rescaled under the theoretical prediction for the critical value $\mu_c$, with the increase in network energy normalized by the square of the inclusion length. We observe an excellent data collapse under this critical value, suggesting that the scaling arguments discussed capture the physics of this transition. In all cases, the bending rigidity of the inclusion is taken to be sufficiently large to avoid any significant deflections. Results are averaged over $60$ realisations of the base network.}
		\label{f7}
	\end{figure*}

	We first consider the increase in energy from a fibre with varying stretching modulus $\mu_i$ in Voronoi architectures. We treat inclusions as stiff in bending, so that fibre deflections are minimal. If this fibre is compliant in extension, so that under the motion of the base network the majority of the change in energy is stored in inclusion stretching, we expect the energy increase to scale as:
	\begin{align}
	\delta\mathcal{H} &\sim \frac{1}{2} \mu_i \epsilon_i^2 L_i \ ,\\
	&\sim \mu_i \gamma^2 \sin^22\theta_i L_i \ ,
	\end{align}
	\noindent where $\epsilon_i \sim \frac{\gamma}{2}\sin2\theta_i$ denotes the affine strain in the inclusion. Now, recall that the transverse spring constant for the fibres is given by $3\kappa/l^3$, for segment length $l$. 
	For random Voronoi networks neighbouring segments are expected to be independent, such that correlations between nearby edges are absent. Then, we assume that we can discuss the system in terms of a single transverse spring constant $k_\perp \simeq \kappa/l_c^3$ by viewing the network as springs in series, similar to elsewhere in the literature \citep{heussinger2006stiff}. If we consider the network energy increase from a stiff fibre (large $\mu_i$), inducing displacements in the $L_i/l_c$ crossing fibres according to a corrected strain field in the axial direction of the fibre, we have:
	\begin{align}
	\left\langle \delta\mathcal{H} \right\rangle &\sim  \frac{L_i}{l_c} \frac{\kappa_b}{l_c^3} \frac{1}{L_i} \int_{0}^{L_i/2}(z \sin2\theta_i\gamma)^2 \diff z  \ , \label{eq_mucrit} \\
	& \sim \frac{\kappa_b\gamma^2\sin^22\theta_iL_i^3}{l_c^4} \ ,
	\end{align}
	\noindent where we have averaged over $z$, which measures the distance along the inclusion fibre from the midpoint. We predict that there exists a transition between these two regimes as $\mu_i$ increases. In this case, the two competing energy contributions must balance at the transition, such that for the critical value $\mu_c$ of the stretching stiffness, we have:
	\begin{align}
	\mu_c\gamma^2\sin^22\theta_iL_i \simeq  \frac{\kappa_b\gamma^2\sin^22\theta_iL_i^3}{l_c^4} \ ,
	\end{align}
	\noindent so that:
	\begin{align}
	\mu_c \approx \kappa_b \frac{L_i^2}{l_c^4} \ .
	\end{align}
	To investigate this hypothesis for the critical value $\mu_c$, we quantify the energy change in Voronoi networks for a variety of simulation parameters in Fig.~\ref{f7}. The left-hand plot shows the results for $4$ different parameter choices, as $\mu_i$ is varied across many orders of magnitude, normalized by the stretching stiffness of the base polymer $\mu_b$. The predicted transition is clear, with an increase in total network energy with $\mu_i$ of over $2$ orders of magnitude, until a plateau region is reached, where further increases in the inclusion axial stiffness do not lead to additional increases in total network energy. The result of scaling the inclusion stretching stiffness $\mu_i$ with the predicted critical modulus $\mu_c$ derived above is shown in the right-hand figure, where the change in energy is scaled by the square of the inclusion length, as suggested by Fig.~\ref{f6} and Eq.~\ref{eq_Li}. We observe an excellent data collapse across all parameter choices under this critical value for the stretching modulus, evidencing that the transition does occur according to the change in deformation modes discussed above. Interestingly, we note that this critical $\mu_c$ can be small for nonaffine networks. Indeed we can rewrite the expression for the critical value as:
	\begin{align}
	\frac{\mu_c}{\mu_b} \simeq \left(\frac{l_b}{l_c}\right)^2\left( \frac{L_i}{l_c}\right)^2 \ ,
	\end{align}
	\noindent where again the bending length-scale is given in terms of the bending and stretching base rigidities as $l_b =  (\kappa_b / \mu_b)^{1/2}$. For nonaffine networks, the ratio of this quantity to the mean segment length $l_c$ is typically small, such that the critical value for $\mu_c$ can be orders of magnitude smaller than that of the base network. This is to be expected for Voronoi networks, which follow a largely affine displacement field at the length scale of the inclusion.
	
	When considering Mikado networks, the assumption utilized in Voronoi architectures, namely that fibre segments are independent and act as springs in series, breaks down due to long base fibre length $L$, and we must consider the spring constant in more detail. In particular, we follow a similar approach to \citet{heussinger2006floppy}, in the context of floppy modes in fibre networks. Consider again the average energy increase from base fibre deflections induced by the stiff inclusion correcting the affine displacement field. If we average over the base fibre segment length distribution $P(l_s) \sim e^{-l_s / l_c}/l_c$, we have:
	\begin{align}
	\langle \delta\mathcal{H} \rangle&\sim \frac{L_i}{l_c} \int_{l_\text{min}}^{\infty}\frac{2}{L_i} \int_{0}^{L_i/2}\kappa_bP(l_s)\frac{z^2\sin^22\theta\gamma^2}{l_s^3} \diff l_s \diff z \ ,\\
	&\sim \frac{\kappa_b\gamma^2\sin^22\theta_iL_i^3}{l_c^2}\frac{1}{l_\text{min}^2} \ ,
	\end{align}
	\noindent where $l_\text{min}$ is a regularizing length, under which the deflection of the fibre segment becomes too energetically expensive, and the bending energy of this segment is instead relaxed through the translation of an entire secondary base fibre, exciting the floppy modes there instead. This minimum length may be determined by balancing the energy content in a deflected short segment with that in an entire secondary fibre, as:
	\begin{align}
	\frac{\kappa_b d^2}{l_\text{min}^3} \simeq \frac{L}{l_c} \int_{l_\text{min}}^{\infty} \kappa_bP(l_s)\frac{d^2}{l_s^3} \diff l_s \ ,
	\end{align}
	\noindent which suggests a minimum length of $l_\text{min} \simeq l_c^2 / L$, and therefore a corrected scaling in Mikado composites of:
	\begin{align}
	\langle \delta\mathcal{H} \rangle&\sim \frac{\kappa_b\gamma^2\sin^22\theta_iL_i^3L^2}{l_c^6} \ .
	\end{align}
	\noindent Balancing this expression with that for the energy in an affinely stretched fibre produces the expression for the critical stiffness $\mu_c$ in Mikado base networks:
	\begin{align}
	\mu_c \simeq \kappa_b \frac{L_i^2 L^2}{l_c^6} \ .
	\end{align}
	
	\begin{figure*}[t]
		\centering
		\includegraphics[width=0.95\textwidth]{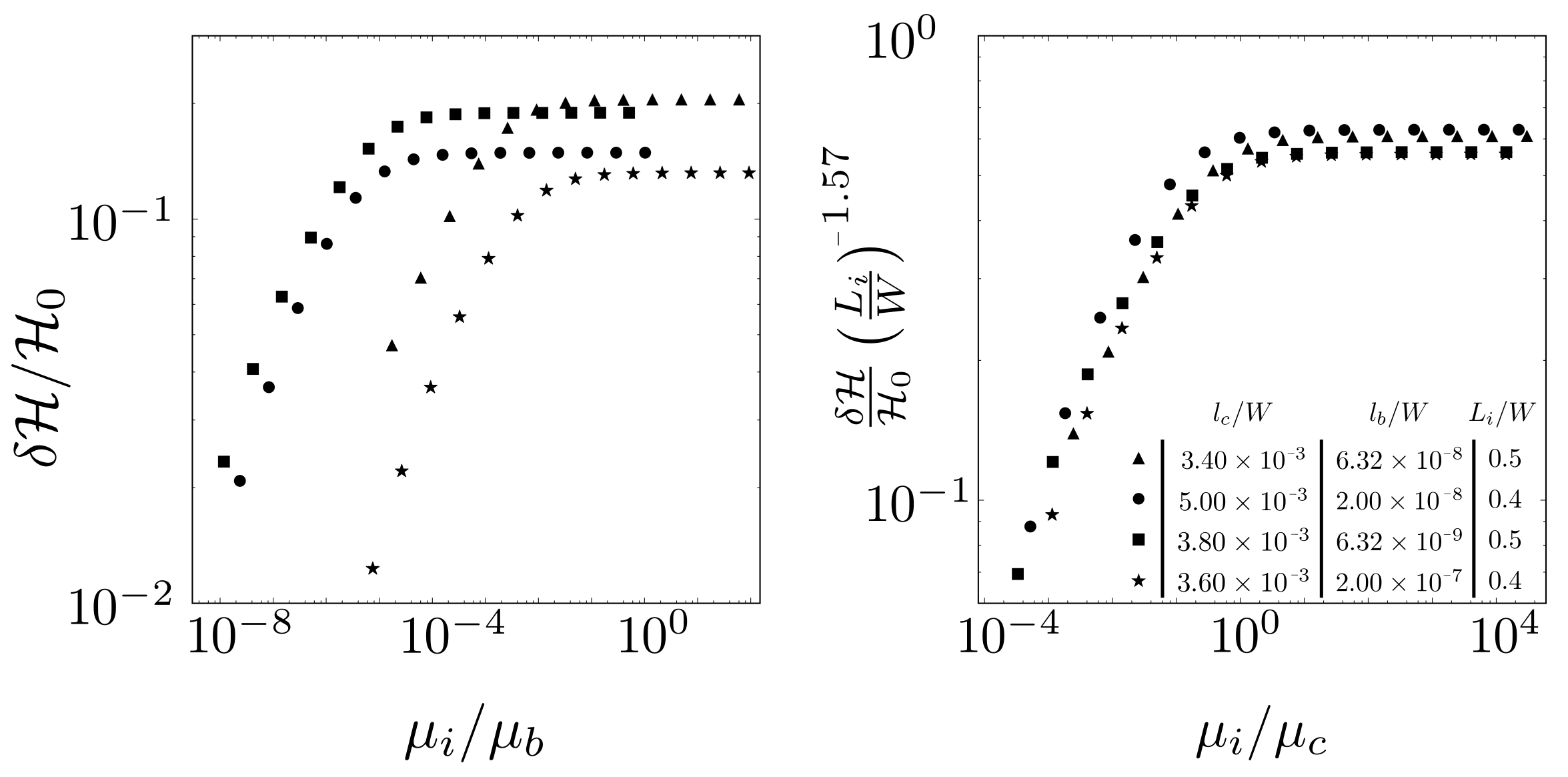}
		\caption{The percentage increase in total network energy for Mikado composites is investigated as the inclusion stretching modulus $\mu_i$ varies. \textbf{(Left)} Similar transitions under varying $\mu_i$ are observed in Mikado composites as in Voronoi cases, shown for four different parameter choices for the base network density and bending stiffness as well as for the inclusion length, normalized by the base fibre rigidity $\mu_b$. \textbf{(Right)} The results are rescaled under the corrected prediction for the critical value $\mu_c$ in Mikado composites. By considering a corrected critical stiffness $\mu_c$ in the presence of additional contributions from Mikado base fibre length, we achieve a data collapse such that the transition for all composites considered occurs under the same length scale $\mu_i/\mu_c$. We note that for equivalent base fibre bending stiffness this transition occurs at larger values of $\mu_i$ than in Voronoi composites. Results are averaged over $60$ realisations of the base network.} 
		\label{f8}
	\end{figure*}
	
	Results for Mikado networks across a variety of parameter choices as $\mu_i$ is varied is shown in Fig.~\ref{f8}. The left hand figure displays the unscaled results, where we again note that a similar transition occurs in Mikado architectures as in Voronoi cases. However, the increase in energy is somewhat diminished, to approximately a $10$ times increase in energy, rather than the $100$ fold increase measured in Voronoi composites. The same data are displayed under the scaled $\mu_i$ in Fig.~\ref{f8}~\textbf{(Right)} where we have again rescaled the change in energy with $L_i$ according to Fig.~\ref{f6} . We find that this corrected critical value again produces a good collapse for the simulation data, evidencing that the corrected value for the critical stiffness captures the network behaviour at the transition. We note that the expression for this critical stretching stiffness is far larger than for Voronoi base networks, owing to the lack of fibre persistence through cross-links.
	
	\begin{figure}[t]
		\centering
		\includegraphics[width=0.48\textwidth]{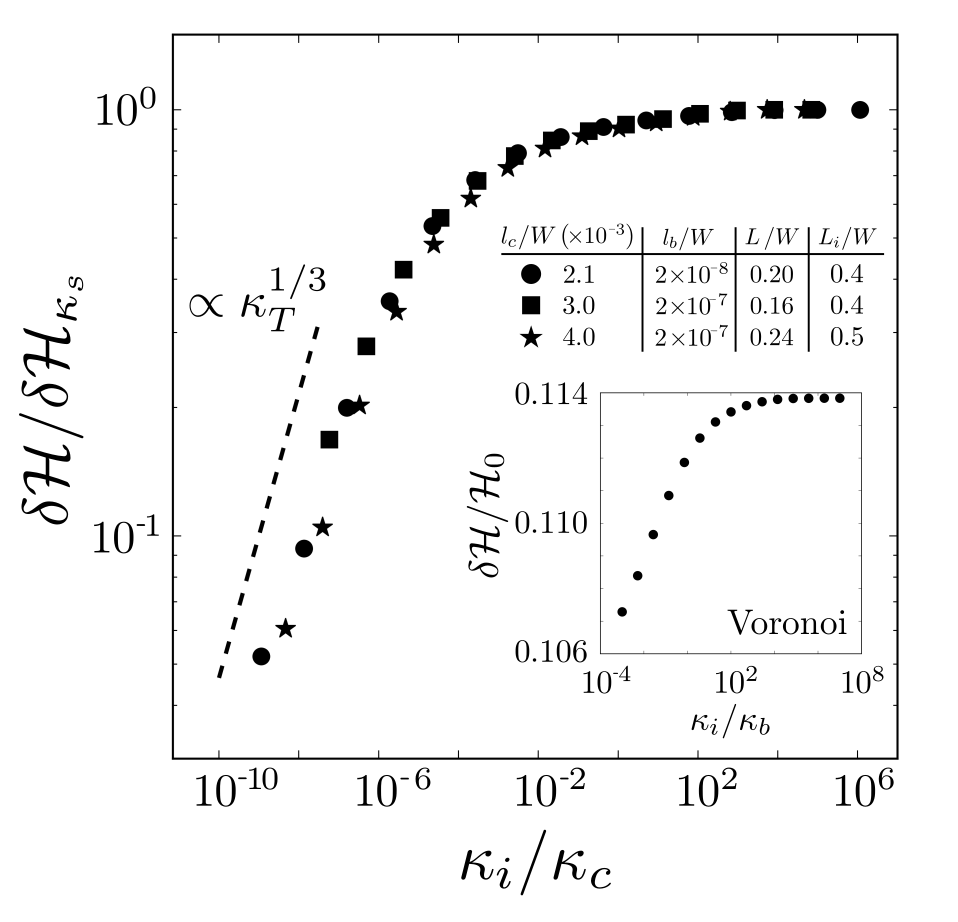}
		\caption{The scaling of the total network energy increase with increasing inclusion bending stiffness $\kappa_i$ is investigated for Mikado composites under three sets of parameter choices within the nonaffine regime. Similar to the results for varying the inclusion stretching modulus, we again observe a transition to a plateaux region, in which further increases to the bending rigidity do not lead to likewise energy increases in the total network energy. The percentage energy increase is scaled according to this plateau value ${\delta\mathcal{H}_\kappa}_s$, while the fibre bending stiffness is scaled according to the critical value $\kappa_c$ predicted from scaling arguments. A good data collapse is observed across the three parameter sets, with the network energy increase scaling close to the $\kappa_i^{1/3}$ predicted. \textbf{(Inset)} The same transition is considered for Voronoi composites. Here, the variation in network energy changes is small (below $1\%$) over $12$ orders of magnitudes in the inclusion stiffness, such that the inclusion bending rigidity is not likely to be the dominant factor in determining the energy increase in Voronoi composites. Results are averaged over $80$ realisations of the base network.}
		\label{f9}
	\end{figure}
	
	Having investigated the transition that occurs as the stretching stiffness of the inclusion increases from compliant to stiff, for high bending rigidity, we now wish to investigate how the network increase in energy varies with the inclusion bending stiffness. Given our discussion of Voronoi base networks, in which displacement nonaffinity is small, we do not expect large deflections in the inclusion, such that varying $\kappa_i$ is not expected to produce significant variations in total network energy. However, the bending stiffness of the inclusion is expected to have important consequences for Mikado base networks, where significant fluctuations in the base network are likely to induce large bending deformations in the inclusion. Consider first a `floppy' inclusion, which deforms according to the background nonaffine Mikado network. The deflections incurred due to the base network are proportional to the size of the nonaffine displacements, scaling as $L$ \citep{heussinger2006floppy}. Then, as the bending stiffness $\kappa_i$ increases from zero, we expect to see an increase in energy according to this fibres deflections. In particular, we can write this energy in the form:
	\begin{align}
	\left\langle \delta\mathcal{H} \right\rangle &\sim \frac{L_i}{l_c} \int_{l_\text{min}^i}^{\infty} \kappa_i P(L_s)\frac{L^2}{L_s^3}\diff L_s \ , \\
	&\sim \kappa_i \frac{L_iL^2}{l_c^2} \left(\frac{1}{l_\text{min}^i}\right)^2 \ , \label{l_min_change}
	\end{align}
	\noindent where $l_\text{min}^i$ is the minimum length for a segment along the inclusion fibre to bend, rather than inducing displacements of crossing fibres, and $L_s$ denotes fibre segment lengths between cross-links along the inclusion. We can determine this length self consistently as follows. For fibre segments possessing this minimum length, energy balance must exist between the deflection of the single segment and the energy in an entire secondary translated base fibre, which, utilizing the expression for $l_\text{min}$ from above can be expressed as:
	\begin{align}
	\kappa_i d^2\left(\frac{1}{l_\text{min}^i}\right)^3 &\sim \frac{L}{l_c}\int_{l_\text{min}}^{\infty}\kappa_b \frac{d^2}{l_s^3}P(l_s)\diff l_s \ , \\
	&\sim \frac{\kappa_bLd^2}{l_c^2}\left(\frac{1}{l_\text{min}}\right)^2 \ , \\
	&\sim \frac{\kappa_bd^2L^3}{l_c^6} \ ,
	\end{align}
	\noindent so that:
	\begin{align}
	l_\text{min}^i &\sim \left(\frac{\kappa_i}{\kappa_b}\right)^{1/3}\frac{l_c^2}{L} , 
	\end{align}
	\noindent and the corresponding increase in energy from Eq.~\ref{l_min_change} is given as:
	\begin{align}
	\left\langle \delta\mathcal{H} \right\rangle &\sim \frac{L_iL^4}{l_c^6}\kappa_i^{1/3}\kappa_b^{2/3} \ .
	\end{align}
	\noindent As such, we expect that the change in energy should scale as $\kappa_i^{1/3}$ in this regime for Mikado composites. Now, as $\kappa_i$ increases, we expect to observe the formation of the low-energy, rigid-body region towards the centre of the inclusion, as deflections in the inclusion become energetically unfavourable. As discussed previously, we anticipate that this leads to additional energy contributions where fibres within the rigid body region interact with distant fibres in the base network, resulting in additional fibre deflections. As the number of fibres crossing the stiff inclusion scales as $L_i/l_c$, and the cross-links along a base fibre as $L/l_c$, we expect that the energy increase here should scale as:
	\begin{align}
	\left\langle \delta\mathcal{H} \right\rangle &\sim \frac{L_iL}{l_c^2}\int_{l_\text{min}}^{\infty} \kappa_b P(l_s) \frac{L^2}{l_s^3}\diff l_s \ , \\
	&\sim \kappa_b\frac{L_iL^5}{l_c^7} \ .
	\end{align}
	\noindent Seeking a balance of the two energy contributions, from the deflections in the inclusion and from the rigid body rotation, leads to a critical value of the inclusion bending rigidity $\kappa_c$ given by:
	\begin{align}
	\kappa_c \simeq \kappa_b\left(\frac{L}{l_c}\right)^3 \ .
	\end{align}
	
	We investigate these hypotheses for inclusions of different bending rigidities by simulating Mikado composites in which inclusions possess high stretching rigidity $\mu_i$ (above the previously identified critical value) and varying $\kappa_i$ over many orders of magnitude. In Fig.~\ref{f9} we illustrate the result of scaling the inclusion bending stiffness by the critical value discussed above for three different parameter choices. The change in energy $\delta\mathcal{H}$ is scaled by the plateau value, denoted $\delta\mathcal{H}_{\kappa_s}$, that is the equivalent quantity for very large $\kappa_i$, after which further increases do not change the result. Under scaling by $\kappa_c$, we observe a good collapse of data across different simulations with various parameters, suggesting that the theory captures the essential physics of the system. In particular, the scaling of the energy change before the plateau region is close to the predicted $\kappa_i^{1/3}$. Inset in Fig.~\ref{f9} is an example analogous transition in Voronoi composites. However, we do not discuss this transition further, as the associated change in energy is small enough to be trivial compared to the increase in energy induced by the axial constraints produced by the inclusion. In fact, the variation of $\kappa_i$ over $12$ orders of magnitude only results in a less than $1\%$ variation in total network energy change $\delta\mathcal{H}/\mathcal{H}_0$. This agrees with our prediction that the smaller nonaffine fluctuations in Voronoi architectures prevent large deflections in the inclusions. These results, for varying inclusions stiffnesses $\mu_i$ and $\kappa_i$ detail an important difference between composites in different architectures. In particular, where filament composites possess long base fibres, we expect that the increase in network energy will depend sensitively on both inclusion stretching and bending rigidities, as in the Mikado composites studied here. However, where the length scales of the base fibres are far smaller than those of the inclusion, and the number of cross-links along each base fibre is low, we expect to see large increases in energy under the introduction of an inclusion. This is predicted even when the inclusion is compliant in both stretching and bending, as seen in the low critical stretching value $\mu_c$, and insensitivity to bending stiffness $\kappa_i$, in Voronoi examples here.
	
	\section{\label{sec:level4}Discussion}
	
	In order gain a better understanding of the mechanics of semiflexible biopolymer assemblies, including cytoskeletal and extracellular matrices, it is necessary to extend the scope of discrete fibre network models to composite materials. Discrete fibre network modelling efforts have begun to move away from simple choices, in which networks are comprised of filaments with identical lengths and mechanical properties. This approach has been successful in reproducing and explaining experimental results \citep{lin2011control}, as well as improving the understanding of theoretical models and the relevant physics \citep{huisman2010semiflexible, van2017criticality}. However, the architectures utilized in such modelling should aim to reflect the diverse possibilities in real composite assemblies, including the topology and geometry of those matrices. In order to investigate the extent to which composite mechanics depend upon the details of the base network architecture and geometry, we investigated two choices for the base, nonaffine matrix. In particular, we investigated Voronoi networks, in which there is no notion of fibre persistence through a network cross-link, and Mikado networks, in which fibres are assumed to be perfectly straight along their backbone. Of course, real biopolymer networks will exist somewhere in between, depending on the thermal and mechanical properties of the fibre matrix under investigation.
	
	The single fibre composites investigated here displayed many differences in their mechanics, owing to the choice of base network architecture. The most significant variations occurred due to dissimilarities between the displacement and rotation nonaffinity they possess. While mechanical regimes are often discussed in terms of the relative energy content in different deformation modes, namely bending and stretching of constituent fibres, we here found that the more relevant definition for affinity was given in terms of deviations in the displacement field from the affine prediction. That is to say, for equivalent energetic definitions of affinity, measured through the ratio of total network energy to the corresponding affine prediction, Voronoi networks exhibit far smaller nonaffine displacement fluctuations than their Mikado counterparts, and it was these fluctuations that lead to distinct responses in the presence of a stiff inclusion. For nonfibrous, Voronoi composites, these departures from affinity were small at the scale of the inclusion, such that the inclusion rotated and translated according to an affine background displacement field. In contrast, the additional length scale $L$ in Mikado composites distorted this picture for the inclusions considered here, with significant nonaffine fluctuations persisting even at long inclusion lengths.
	
	By considering a bulk affine strain, we were able to explain the localisation of energy at the terminations of the inclusion in terms of a correction to a background strain field that was largely affine in the absence of the inclusion. In particular, when a sufficiently stiff inclusion was placed into a base network, rather than stretching according to the base displacements of the compliant network, nonaffine corrections were introduced that deformed the base network in the axial direction of the inclusion. As such, spatially nonaffine regions in the network corresponded to regions of higher energy. We also observed a low energy region towards the centre of the inclusion body, where coincident base fibres underwent rigid body rotations with the inclusion. 
	
	Considering corrections to the affine strain field in the axial direction of the inclusion so as to prevent inclusion stretching allowed a theoretical prediction for the scaling of the increase in network energy with inclusion orientation $\theta_i$ as well as length $L_i$. It was found that, in both cases, the theory agreed well with data from simulations for nonfibrous composites. However, additional energy contributions were observed in Mikado networks, leading to asymmetry in inclusion angle $\theta_i$. We explained this by considering the rigid body region adjacent to the inclusion, where fibres coincident to the inclusion antagonized distant base fibres. This interpretation lead to a correction in the scaling of energy increases in Mikado composites, with the introduction of an additional term proportional to $\sin^4\theta_i$ resulting in a good agreement with synthetic data. It was observed that the scaling with $L_i$ became nontrivial in Mikado cases, where competing contributions from the suppression of nonaffine deformation, rigid body rotations and axial displacement corrections contributed to a lessened dependence on $L_i$ within the range considered here. We predict that for a large enough ratio $L_i/L$ the expected scaling of $L_i^2$ could be recovered as the nonaffine rotations in the base network become insignificant.
	
	While we initially considered inclusions that were far stiffer in both bending and stretching than base fibres, we observed transitions as both of these parameters were varied from low to high values. By considering balances between the energy content of a deformed inclusion and that resulting from displacements induced in the compliant base network by a stiff inclusion, the locations of these transitions were identified for $\mu_i$ and $\kappa_i$ in each network geometry. Interestingly, the critical value for $\mu_i$ in nonfibrous composites occurs at perhaps surprisingly low values, smaller than the corresponding quantity in the base network for many examples considered. The critical value in Mikado composites was greater, owing to the persistence of filaments through network cross-links, which was understood through considering floppy modes along Mikado fibres. The picture for bending rigidity $\kappa_i$ was more complex. For networks deforming approximately affinely at the scale of the inclusion, such as the Voronoi networks considered in this work, the bending rigidity was largely unimportant, leading to only small increases in network energy across many orders of magnitude when compared with the other energy contributions. In contrast, large values of bending rigidity $\kappa_i$ relative to $\kappa_b$ were required in Mikado composites to out-compete the nonaffine displacements occurring in the base network. This provides an important prediction for real fibre network systems, where even quite compliant inclusions placed within homogeneous matrices could lead to large changes in stiffness, while far longer, stiffer inclusions might be required to induce similar stiffness increases in matrices possessing long, straight polymer lengths. For both Voronoi and Mikado composite types, the collapse of data from computer simulations under our predictions evidenced the validity of our theory, such that the mechanics of low density double-fibre networks can be understood across a broad range of parameters throughout the nonaffine regime.
	
	While the results presented here discuss a range of situations for both the base polymer system and the inclusion, we note that real world networks have additional complexities that cannot be addressed within this two-dimensional model. In particular, when extended to three dimensions, additional deformation modes are available to the compliant base network; it is not clear how the addition of local twisting and torsional effects will alter the results presented within this work. However, as fibres in three-dimensional environments might be expected to be less affine, owing to a larger isostatic threshold and additional bending degrees of freedom, it could suggest that the behaviours observed here in 2D are present even in stiffer polymers in three-dimensions. In contrast, it could also be the case that increased nonaffinity allows the base matrix to accommodate the constraints of the stiff inclusion in local nonaffine rotations, similar to the Mikado composites discussed in this work for small inclusion lengths $L_i$. 
	
	While not discussed in this work, we note that another example of a composite fibre network system is found in matrices in which mechanically active cells are present. We might also expect that polarized, contractile cells, which are able to deform the surrounding matrix, produce similar mechanical phenomena to those discussed in this work. Such cells, which can respond to the stiffness of the substrate, are likely to experience similar substrate displacements to the stiff inclusions modelled here. In mechanically active cells, we expect to observe contractile activity along the cell long axis, whence the theory discussed here for stiff inclusions can be applied. This approach could allow for predictions in cellularized fibre network systems, where other relevant phenomena include cell reorientation and alignment with induced strain direction, and sensitivity to substrate fibre bending stiffness.
	
	\section{Conclusions \& summary}
	
	We have investigated the mechanics of composites formed when a single inclusion is placed within fibre networks constructed from two different architectures by subjecting such composites to small shear deformations. To investigate how the mechanics of these systems can deviate from that observed in continua, fibre networks within the bending-dominated regime were used throughout. By investigating a measure for displacement nonaffinity, we found that even when networks are located within a similarly bending-dominated regime, the nonaffine strain fields can differ by orders of magnitude. By considering spatial measures of displacement nonaffinity in relation to the energy content in the compliant matrix, we observed that nonaffine displacements correlate with larger base fibre deformations. We interpreted the stiff inclusion as imposing a correction to the affine displacement prediction in the axial direction of the inclusion, and produced theoretical predictions for the scaling of the network energy. In nonfibrous Voronoi examples, simulation data agreed closely with theory for both inclusion orientation and length. However, additional deformations were induced in Mikado composites, due to the additional length scale of the base fibres. By considering additional contributions to network energy owing to the existence of a rigid body region adjacent to the inclusion, we recovered a good agreement between data and theory for inclusion orientation in such composites. We observed that bending stiffness was largely unimportant in nonfibrous Voronoi composites, yet relatively low inclusion stretching stiffnesses could produce large increases in the total network energy. In Mikado composites, the critical values for the moduli were found to be greater than in the nonfibrous examples, as the additional nonaffinity in the matrix lead to more complex deformation behaviours. Finally, theory predicted the existence and location of transitions for the network deformations as the inclusion stretching and bending stiffnesses increased, producing a convincing data collapse across multiple parameter choices. In turn, this enables a simple, evidence-based understanding of the mechanics of networks with a single fibre inclusion, valid across a large region of parameter space.
	
	\begin{acknowledgments}
	DLH acknowledges funding from the Engineering and Physical Sciences Research Council (EPSRC) [grant number EP/G03706X/1]. JAG acknowledges funding from the Computational Horizons in Cancer (CHIC) project under EU FP7 [grant agreement number 600841]. In compliance with EPSRC's open access initiative, the data in this paper is available from http://dx.doi.org/xxx/xxx.
	\end{acknowledgments}
	
	\bibliographystyle{apsrev4-1}
	\bibliography{master_bib}
	
\end{document}